\documentclass[
  prb,
  twocolumn,
  superscriptaddress,
  showpacs]
  {revtex4}

\usepackage{graphicx}

\newcommand{\e}{\text{e}}
\newcommand{\Tc}{T_{\text{c}}}
\newcommand{\betac}{\beta_{\text{c}}}

\begin{document}

\title{Finite-temperature order-disorder phase transition in a \\
cluster model of decagonal tilings}

\author{Michael Reichert}
\altaffiliation[Present address: ]{Fachbereich Physik, Universit\"at Konstanz,
D-78457 Konstanz, Germany}
\email{michael.reichert@uni-konstanz.de}
\affiliation{Institut f\"ur Theoretische und Angewandte Physik,
Universit\"at Stuttgart, \\ D-70550 Stuttgart, Germany}
\affiliation{Fachbereich Physik, Universit\"at Konstanz,
D-78457 Konstanz, Germany}
\author{Franz G\"ahler}
\affiliation{Institut f\"ur Theoretische und Angewandte Physik,
Universit\"at Stuttgart, \\ D-70550 Stuttgart, Germany}

\date{\today}

\begin{abstract}
In a recent paper [{\it Cluster Model of Decagonal Tilings} 
(to be published in Phys.~Rev.~B)], we have introduced a cluster model for
decagonal tilings in two dimensions. This model is now extended
to three dimensions. Two-dimensional tilings are stacked on top
of each other, with a suitable coupling between adjacent layers.
An energy model with interactions leading to a perfect decagonal 
quasicrystal at low temperatures is studied by Monte Carlo simulations. 
An order parameter is defined, and its dependence on temperature 
and system size is investigated. Evidence for a finite-temperature 
order-disorder phase transition is presented. The critical
exponents of this transition are determined by finite-size scaling.
\end{abstract}


\maketitle

\section{Introduction}
\label{sec_intro}

It is well known\cite{kal89} that, in two dimensions (2D), perfect 
quasicrystalline order cannot be stable at positive temperature
if the interactions have finite range. At positive temperature, 
a 2D quasicrystal is always in the random tiling phase without 
long-range order, the ``transition'' from the ordered to the 
disordered state being at $T=0$. For three-dimensional (3D) 
quasicrystals, on the other hand, this order-disorder phase 
transition is expected to occur at positive temperature.\cite{hen91} 

The low-temperature state is also called the {\it locked} 
phase,\cite{hen91} because its phason degrees of freedom 
are frozen (locked). The high-temperature state is accordingly 
called {\it unlocked} phase. Here, the thermal energy 
is sufficiently high to excite the phason degrees of freedom.

3D axial quasicrystals, in particular decagonal ones, can usually
be regarded as periodic stackings of 2D layers, each of which 
is quasiperiodic. Geometrically, these layers can be described as 
decorations of quasiperiodic tilings like the Penrose tilings.
Henley\cite{hen91} has proposed to model axial quasicrystals as
stackings of 2D tilings with a suitable coupling between adjacent 
layers, in addition to the coupling inside the 2D layers. Such 
layered tiling models built up from 2D Penrose tilings have been 
studied by Jeong and Steinhardt,\cite{jeo93} who indeed found
a finite-temperature order-disorder phase transition. 

In a recent paper\cite{part1} (hereafter referred to as Paper~I), 
we have introduced a cluster model for 2D decagonal tilings.
In the present article, we will now extend this model to 3D stackings.
Our aim is to investigate an order-disorder phase transition, too, but 
this time in the framework of {\it cluster coverings}.

In Sec.~\ref{sec_2d-model}, we first give a brief summary of Paper~I,
explaining the important features of our cluster model. This includes 
the principle of cluster density maximization and the ordering of 
random structures to perfect ones by a coupling between overlapping 
clusters. In Sec.~\ref{sec_3d-model}, we extend this model to 3D by
stacking the 2D layers on top of each other. An additional coupling
between adjacent layers is introduced. For this 3D model we define 
in Sec.~\ref{sec_orderparam} an order parameter to distinguish 
between the ordered quasicrystalline phase at low temperatures and 
the disordered random tiling phase at high temperatures. By means of 
this order parameter, the 3D system can then be investigated for a 
potential order-disorder phase transition, which is
done in Sec.~\ref{sec_phasetrans} using Monte Carlo (MC) simulations.

\section{Two-dimensional cluster model \protect\\ (Brief summary)}
\label{sec_2d-model}

\begin{figure}[b]
\includegraphics[width=\columnwidth]{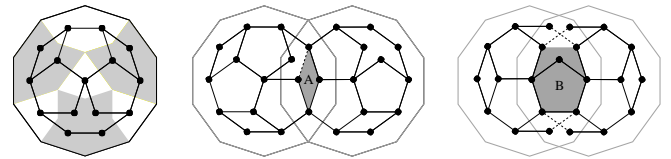}
\caption{Vertex cluster, superimposed on the Gummelt decagon (left),
representative A-overlap (middle) and B-overlap (right). This cluster
enforces the relaxed overlap rule.}
\label{fig_cluster}
\end{figure}

In Paper~I, different versions of overlap rules for clusters in a 
2D covering model of the Penrose pentagon tiling (PPT) are discussed.
The first version, which has a very natural realization in terms 
of a vertex cluster in the PPT (Fig.~\ref{fig_cluster}), is called 
the {\it relaxed} rule. If we compare this cluster with the 
well known Gummelt decagon,\cite{gum96,gum00,gum02} we see that it 
enforces the correct orientation of the large B-overlaps, but not 
the orientation of the smaller A-overlaps (Fig.~\ref{fig_cluster}).

As an alternative to {\it cluster covering} as an ordering principle 
for quasicrystals, the principle of {\it cluster density maximization}
is considered as well. A statistical model is built where each
random tiling is assigned an energy which is just the negative of the
number of vertex clusters, thus mimicking the cohesion energy
of the clusters. 

\begin{figure}[t]
\centerline{\includegraphics[width=4.5cm]{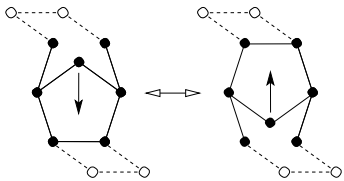}}
\caption{Representative example of a flip move in the PPT. (The
complete set of possible flip configurations is shown in Paper~I.)}
\label{fig_flip}
\end{figure}

The density of clusters can then be maximized in a MC
simulation by simulated annealing, using flip moves like the one shown in 
Fig.~\ref{fig_flip}. The simulation algorithm was a usual Metropolis 
importance sampling scheme:\cite{new99,tan90} A proposed 
flip is accepted with probability $p=1$ if it decreases the energy 
($\Delta E<0$), but only with probability $p=\e^{-\beta\Delta E}$ 
if the energy is increased ($\Delta E>0$). 

\begin{figure}[b]
\includegraphics[clip,width=\columnwidth]{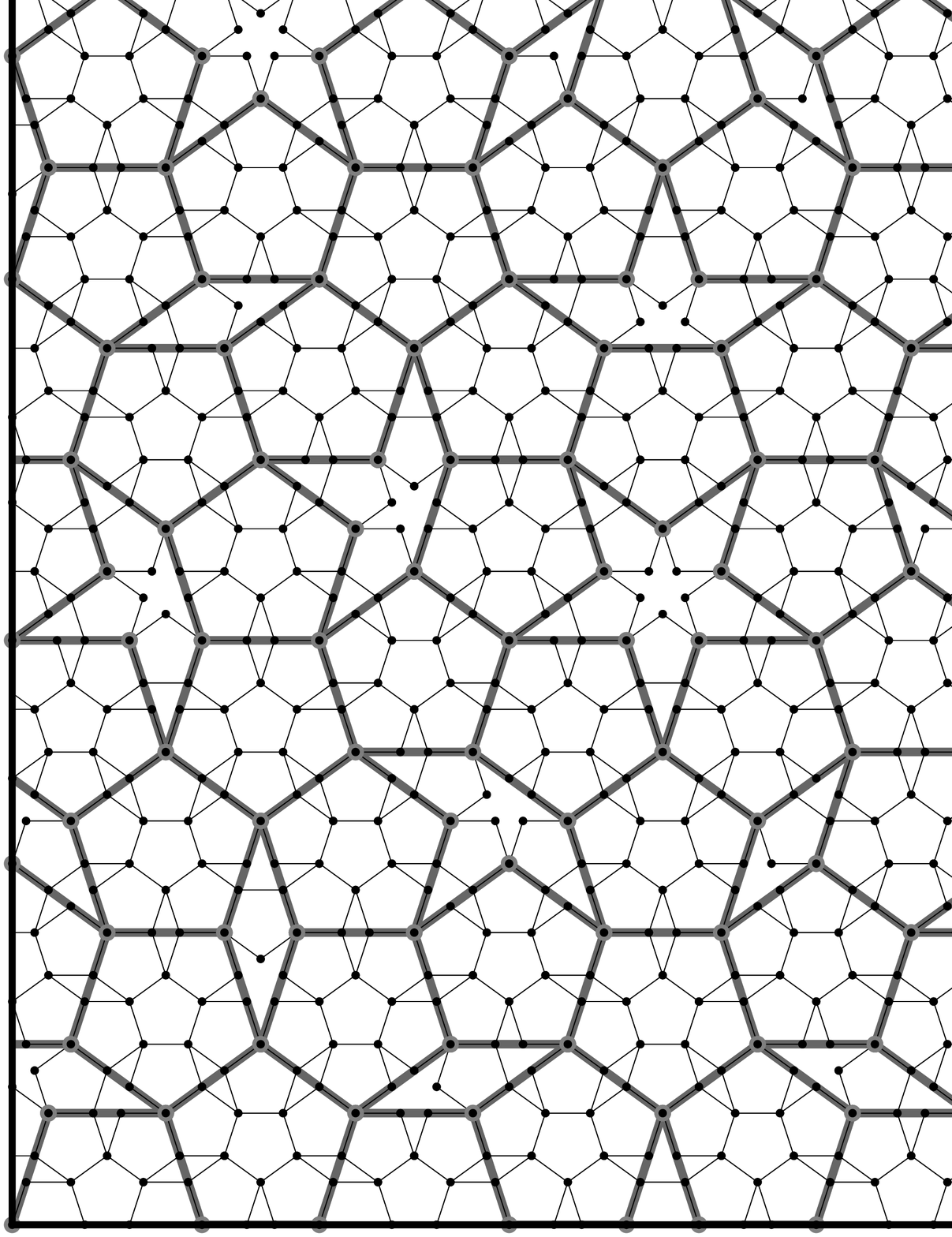}
\caption{Structure with maximal cluster density. The cluster centers
form the vertices of a supertile random PPT (gray lines).}
\label{fig_supertiling}
\end{figure}

The states of maximal cluster density are supertile random PPTs 
(Fig.~\ref{fig_supertiling})
with an edge length $\tau^2$ times that of the underlying tiling (where
$\tau=(1+\sqrt{5})/2$ is the golden number). 
Naturally, they contain also cluster overlaps which do not satisfy the
constraints of Gummelt's {\it perfect} overlap rule, since the {\it relaxed} 
rule does allow for disoriented A-overlaps because of the 
missing interior orientation of the rhombus (Fig.~\ref{fig_cluster}). 

\begin{figure}[t]
\includegraphics[width=\columnwidth]{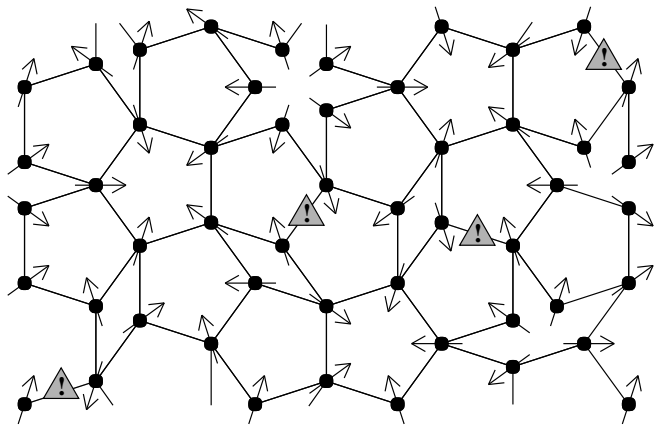}
\caption{Supertiling with cluster orientations indicated by arrows.
The forbidden A-overlaps, corresponding to tile edges with antiparallel
arrows, are marked.}
\label{fig_arrow-model}
\end{figure}

In a second step, a {\it coupling between overlapping clusters} is
added which penalizes such defects. To keep the cluster density 
constant, the simulations are run at the level of the supertiling. 
Each cluster is represented by a vertex plus an arrow indicating 
the orientation of the cluster (Fig.~\ref{fig_arrow-model}). 

\begin{figure}[t]
\includegraphics[width=\columnwidth]{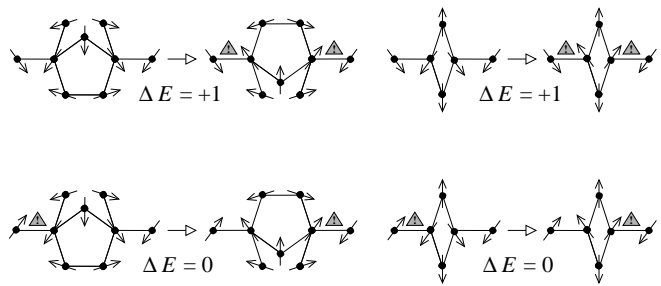}
\caption{Creation (top) and shift of defects (bottom) by single
flips of the two types: hexagon flip (left) and rhombus flip (right).}
\label{fig_defect-moves_xy}
\end{figure}

As MC moves we can still use hexagon flips like the ones in 
Fig.~\ref{fig_flip}, provided that the orientations of neighboring 
clusters are updated consistently with the underlying tiling 
(Fig.~\ref{fig_defect-moves_xy}, left). Additionally, on the level of
the supertiling there is a new type of flip: The rhombus flip
(Fig.~\ref{fig_defect-moves_xy}, right) only 
changes the orientations of the clusters on the obtuse rhombus corners, 
but keeps the tiling itself fixed. In comparison with the basic flip 
moves in the Penrose rhombus tiling\cite{hen91,tan90} (which were used 
in the simulations of Jeong and Steinhardt \cite{jeo93}), the hexagon  
flip corresponds to D-type configurations and the rhombus flip to Q-type
configurations of the Penrose rhombus tiling.\cite{bru81}  

As can be seen from Fig.~\ref{fig_defect-moves_xy}, it is possible to 
create, annihilate, or shift defects by these flips. The defects 
correspond to tile edges with antiparallel arrows at their ends (marked 
in Fig.~\ref{fig_defect-moves_xy}). Hence, we assign a positive energy 
to the creation of a new defect and use this energy model in a 
Metropolis MC scheme. 

By simulated annealing it is shown that the cluster coupling model 
is capable of ordering the supertile random PPTs to perfectly 
quasiperiodic structures at low temperatures. Since, in our 3D 
simulations, we want to study transitions between perfect order 
and disorder, we will use this second kind of model for the layers
in our stacking of coverings.

\section{Extension to three dimensions}
\label{sec_3d-model}

We will now consider 3D stackings of our 2D covering model. The 
{\it intra}layer interactions are those discussed above, which favor
Gummelt's overlap rule, and hence produce perfectly ordered 
structures inside a layer at low temperatures. In addition, a new 
coupling {\it between} the layers has to be introduced. This
{\it inter}layer coupling is chosen such that it prefers adjacent
layers to be congruent. Then the ground state consists of identical,
perfectly orderd layers. In the high-temperature state, the individual
layers are random tilings which need not to be congruent, so that there 
is also disorder in the stacking direction.

\begin{figure}[t]
\centerline{\includegraphics[width=4.5cm]{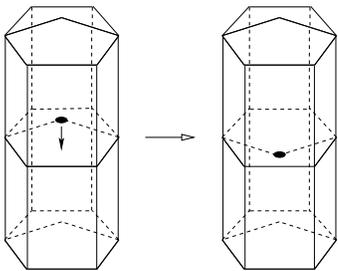}}
\caption{Flip constraint for the interlayer coupling. The flip of the
interior vertex of the middle-layer hexagon violates the congruence of
the layers and hence costs energy.}
\label{fig_flip-constraint}
\end{figure}

Following an idea of Henley\cite{hen91} and analogously to Jeong
and Steinhardt,\cite{jeo93} we formulate a {\it flip constraint} 
for the {\it inter}layer coupling. The vertex inside a hexagon in a certain 
layer can be flipped only if there is also a hexagon with coinciding 
boundary (ignoring the interior vertex) in the adjacent layers above
and below (Fig.~\ref{fig_flip-constraint}). An analogous rule is 
imposed for rhombus flips: There must be a coinciding rhombus above 
and below in order that the middle rhombus can be flipped. These 
constraints maintain a minimal correlation between adjacent layers.

\begin{figure}[t]
\centerline{\includegraphics[width=\columnwidth]{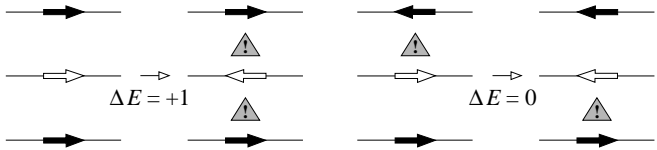}}
\caption{Creation (left) and shift of interlayer defects (right). The
arrows represent here schematically the orientations of hexagons or rhombi,
respectively, in the layers.}
\label{fig_defect-moves_z}
\end{figure}

As we want to favor congruent layers in the ground state, energies
are assigned to any mismatch between adjacent layers. A flip
can then either create or remove two mismatches, or it can shift a 
mismatch up- or downwards (Fig.~\ref{fig_defect-moves_z}).

\section{Definition of an order parameter}
\label{sec_orderparam}

In order to analyse the transition from the disordered phase at high
temperatures to the ordered phase at low temperatures, a suitable order 
parameter has to be defined, which can distinguish between perfect order 
and disorder.\cite{jeo93,dot94}

\begin{figure}[b]
\centerline{\includegraphics[width=\columnwidth]{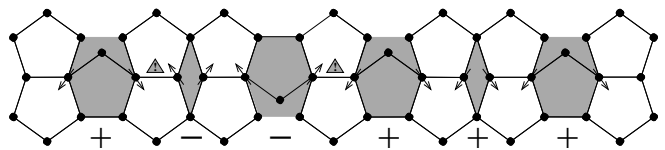}}
\caption{Worm line consisting of hexagons and rhombi. Each hexagon
and rhombus is assigned a ``spin variable'' $+$ or $-$. The presence 
of defects leads to an alternation of the value of this variable along 
the line.}
\label{fig_spin-chain}
\end{figure}

Within one layer, perfect order means the absence of disoriented 
A-overlaps, as discussed in Sec.~\ref{sec_2d-model}. In a continuous
sequence of hexagons and rhombi along a straight line 
(Fig.~\ref{fig_spin-chain}), the orientations are then all the same,
whereas a disoriented A-overlap switches to the opposite orientation.
These lines correspond to the so called ``worms'' or ``Ammann lines'' 
in the Penrose rhombus tiling,\cite{soc86,pav89} where again the 
rhombus in the PPT corresponds to Q-type configurations of Penrose 
rhombi, and the hexagon to D-type configurations. Flipping a hexagon 
or a rhombus in a perfect worm creates mismatches and interrupts the 
sequence of equal orientations along the worm line.

Therefore, if we characterize the orientation of each hexagon 
and rhombus along a straight line by a variable $s_i=\pm$
(Fig.~\ref{fig_spin-chain}), we can compare the worm line to a
1D ``spin chain'', assigning ``spin up'' for one orientation of 
hexagons and rhombi and ``spin down'' for the other. If we sum up 
all the spin variables $s_i$ along a line, they will average out 
to zero in the random tiling phase. In the ordered phase, all the 
spins have the same orientation, so that the sum is proportional
to the length of the chain.

\begin{figure}[t]
\centerline{\includegraphics[width=\columnwidth]{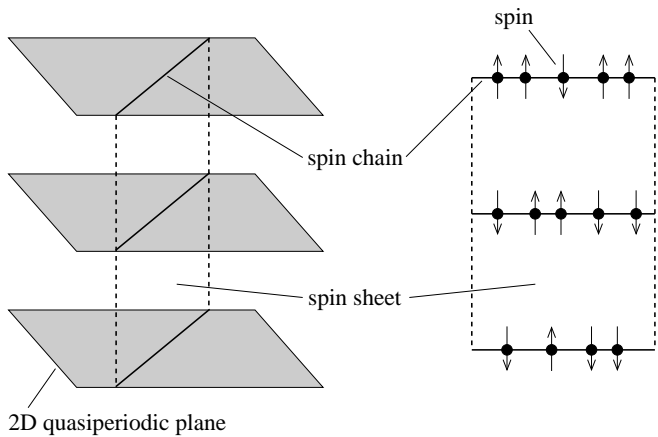}}
\caption{``Spin chains'' one atop the other are combined to a ``spin
sheet''.}
\label{fig_spin-sheet}
\end{figure}

We extend this picture to 3D by combining spin chains atop each other 
to ``spin sheets'' (Fig.~\ref{fig_spin-sheet}), comparable to a 2D 
spin system. In the ordererd phase, all the layers are congruent, 
and thus all the spins within one sheet have the same orientation. 
In the disordered phase, the sum over all spins in a sheet will again 
be zero. However, even in the perfectly ordered phase, parallel spin 
sheets do not necessarily have the same spin orientations. Therefore, 
absolute values have to be taken for each sheet separately. The
{\it order parameter} (``magnetization'') thus becomes 
\begin{equation}
M=\frac{1}{n}\sum\limits_{\{\mathcal{S}_j\}}
\bigg|\sum\limits_{i\in\mathcal{S}_j}s_i\bigg|
\,,
\end{equation}
where $n$ is the number of spins (hexagons and rhombi) considered.
The inner sum, of which the absolute value is taken, runs over all 
hexagons and rhombi (spins $s_i$) lying in the $j$th spin sheet 
$\mathcal{S}_j$. Afterwards, the contributions of all 
parallel spin sheets $\mathcal{S}_j$ are added. Such an order 
parameter is defined separately for each of the five directions 
in the tiling. As we have to use periodic approximants with periodic 
boundary conditions, some spin sheets may wrap around the torus 
several times, which has to be taken into account when determining 
the index of the spin sheet on which a given spin is located.

The value of this order parameter is one in the perfectly ordered 
case, since then all the spins in a sheet have the same orientation.
In the totally disordered phase, the spins average out to zero already
along the chains, so that the order parameter is zero. The magnetization
$M$ therefore provides a well defined and suitable order parameter
for the detection of perfect quasiperiodic order.

In addition to the magnetization, we also define the corresponding 
``susceptibility'' $\chi$, which measures the {\it fluctuations} of 
the order parameter. In analogy to the magnetic susceptibility, we set
\begin{equation}
\chi=n\beta\left(\langle M^2\rangle-\langle M\rangle^2\right) \,,
\end{equation}
where $\beta$ is the inverse temperature.

\section{Order-disorder phase transition}
\label{sec_phasetrans}

The behavior of the order parameter has been investigated by MC
simulations at different temperatures, using a series of stacked
periodic approximants. In order to work out real 3D effects, 
the number of layers is proportional to the linear dimension of the
approximant. We used approximants of order $f_{k+1}/f_{k}$ in the
$x$-direction, and of order $f_{k}/f_{k-1}$ in the $y$-direction.
which have the least number of defects.\cite{ent88} 
$f_{k}$ is the $k$th Fibonacci number ($f_{0}=0$, $f_{1}=1$, 
$f_{k+1}=f_{k}+f_{k-1}$). The linear dimension is then of the
order of $f_{k-1}$, so that we took $f_{k-1}$ layers in the
stacking direction. This resulted in systems with
total vertex numbers $N$ = 141 ($k=5$), 615 (6), 2576 (7), 10959 (8), 
and 46347 (9). The run lengths were of the order of some 100,000 MC
sweeps at each temperature, where perfoming one MC sweep means to
choose $N$-times a vertex randomly. The correlation time was found
to be short compared to the run length, although no precise 
measurement has been made.

\begin{figure}[t]
\includegraphics[width=\columnwidth]{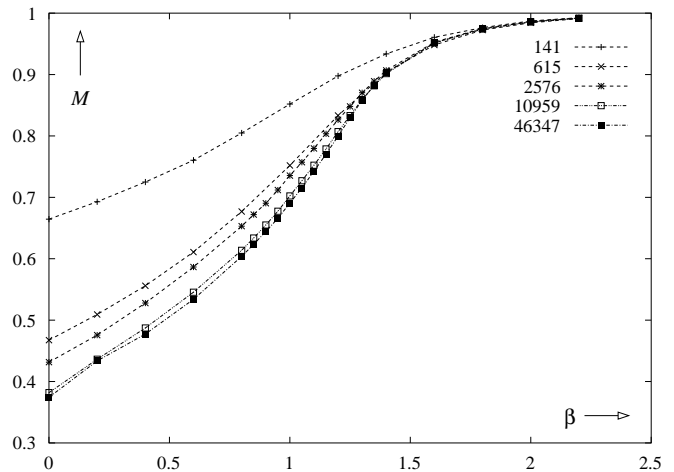}
\caption{System size dependence of the sheet magnetization vs.~inverse
temperature.}
\label{fig_magnet}
\end{figure}

\begin{figure}[t]
\includegraphics[width=\columnwidth]{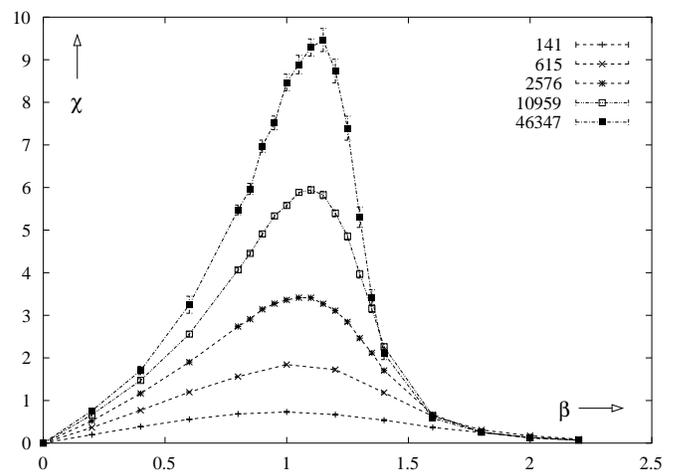}
\caption{System size dependence of the susceptibility vs.~inverse
temperature.}
\label{fig_suscept}
\end{figure}

A Metropolis importance sampling scheme\cite{jeo93,new99,tan90} is 
used for the simulations, similar to the one used in Paper~I for the 2D 
simulations. For each proposed MC move, the total energy change 
$\Delta E$ due to intralayer {\it and} interlayer couplings is computed. 
The move is then accepted with probability $p=1$ if the energy 
is decreased and with probability $p=\e^{-\beta\Delta E}$ if the energy 
is increased. The results for the magnetization and the susceptibility 
are shown in Figs.~\ref{fig_magnet} and \ref{fig_suscept}, respectively, 
for one representative direction in the tiling. 

Fig.~\ref{fig_magnet}
illustrates how the magnetization curves converge to $M=1$ at zero
temperature ($\beta\to\infty$). Furthermore, the magnetization at a
fixed temperature above the critical point ($\beta<\betac$, with
$\betac\approx1.2$) decreases with increasing system size.
The magnetization curves should tend to $M=0$ at temperatures above the 
critical point, but they do so only very slowly. This can be understood 
as follows. In the random tiling phase, many spin sheets contain only 
very few spins, so that on those sheets the spins do not completely 
average out to zero. If for the magnetization only sheets are taken into
account which contain a number of spins sufficiently high for a good 
statistics within the sheet, the value of $M$ can be decreased considerably.
At $\beta=0$, $M$ decreases for the system with 10959 vertices from
$M\approx 0.4$ down to $M\approx 0.1$, and even further down to 
$M\approx 0.08$ for a larger system with about 75000 vertices. 

\begin{figure}[t]
\includegraphics[width=\columnwidth]{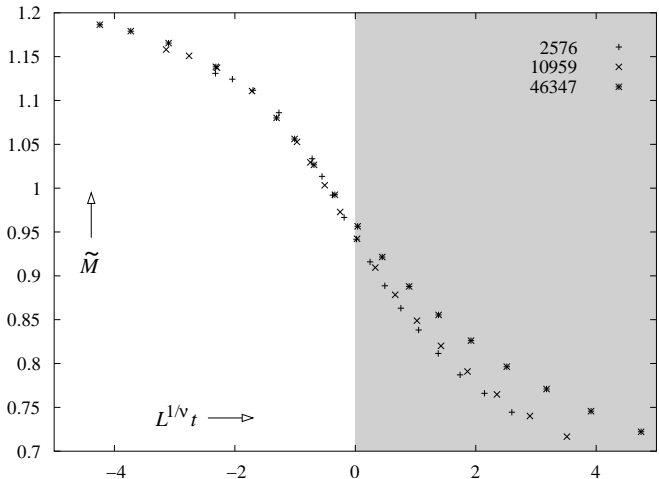}
\caption{Finite-size scaling plots $L^{\beta/\nu}M$
vs.~$L^{1/\nu}t$ for the magnetization with the exponents $\beta=0.08$
and $\nu=1.6$. The critical exponent $\beta$ and hence the scaling
function $\tilde{M}$ are only defined below 
the critical temperature, i.~e., for the data collapse only the range
$t\le 0$ has to be considered.}
\label{fig_scaling_magnet}
\end{figure}

\begin{figure}[t]
\includegraphics[width=\columnwidth]{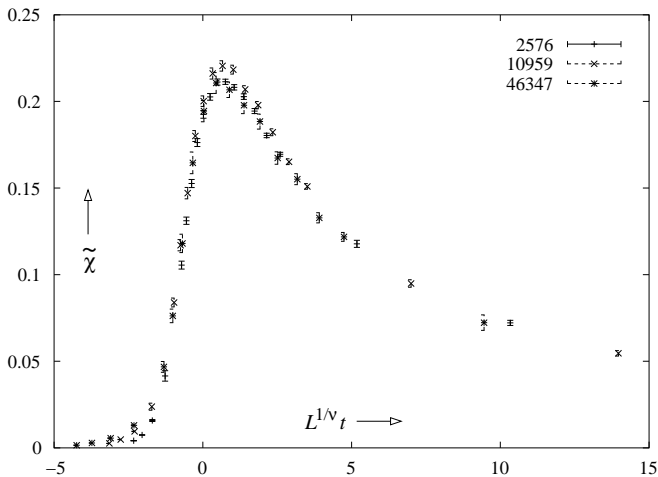}
\caption{Finite-size scaling plots $L^{-\gamma/\nu}\chi$
vs.~$L^{1/\nu}t$ for the susceptibility with the exponents
$\gamma=1.7$ and $\nu=1.6$.} 
\label{fig_scaling_suscept}
\end{figure}

The magnitude of the susceptibility, shown in Fig.~\ref{fig_suscept},
seems to diverge close to the transition temperature with increasing
system size, which is another evidence for a phase transition. 
The maximum of the susceptibility yields a critical temperature of 
$\betac\approx 1.2$ in the limit of infinite system size.

Both the behavior of the magnetization and the susceptibiliy are
consistent with a temperature-driven transition between the
ordered (locked) and the disordered (unlocked) phase. To obtain the
critical exponents of this phase transition, the method of
finite-size scaling\cite{new99} has been used to fit the measured 
curves for magnetization $M_L(t)$ and susceptibility $\chi_L(t)$
to the scaling functions
\begin{eqnarray}
\label{eq_scaling_M}
\tilde M(L^{1/\nu}t) & = & L^{\beta/\nu}M_L(t) \quad 
(\text{for}\hspace{1ex}t\le 0)\,, \\
\label{eq_scaling_chi}
\tilde\chi(L^{1/\nu}t) & = & L^{-\gamma/\nu}\chi_L(t) \,,
\end{eqnarray}
where $L$ is a characteristic length of the system (we use $L=N^{1/3}$) 
and $t=T/\Tc-1$ the reduced temperature (with $\Tc=1/\betac$). 
The critical exponents
$\beta$ and $\gamma$ are defined by the behavior of
magnetization and susceptiblity close to the transition temperature:
\begin{eqnarray}
M & \propto & |T-\Tc|^\beta \quad (\text{for}\hspace{1ex}T\le\Tc) \,, \\
\chi & \propto & |T-\Tc|^{-\gamma} \,.
\end{eqnarray}
The critical exponent $\nu$, which determines the behavior of the
correlation length, $\xi\propto|T-\Tc|^{-\nu}$, has not been measured 
explicitly. By fitting the parameters $\beta$, $\gamma$, and
$\nu$ in Eqs.~(\ref{eq_scaling_M}) and (\ref{eq_scaling_chi}) such
that all the curves superimpose (Figs.~\ref{fig_scaling_magnet} and
\ref{fig_scaling_suscept}), we obtain the critical exponents
\begin{eqnarray}
\beta & \approx & 0.08 \,, \\
\gamma & \approx & 1.7 \,, \\
\nu & \approx & 1.6 \,.
\end{eqnarray}
The uncertainty of these values is about 10\% for $\gamma$ and $\nu$. 
In the case of $\beta$, the spread of values giving a reasonable fit
is even larger.

\begin{figure}[b]
\includegraphics[width=\columnwidth]{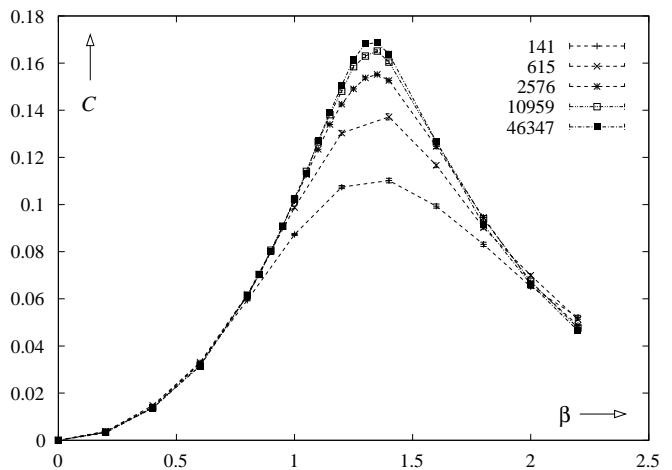}
\caption{System size dependence of the specific heat vs.~inverse
temperature.} 
\label{fig_spec_heat}
\end{figure}

We have also measured the energy $E$ and the specific heat 
$C=N^{-1}\beta\left(\langle E^2\rangle-\langle E\rangle^2\right)$, 
where $N$ is the number of vertices and $\beta$ 
the inverse temperature. The result is shown in Fig.~\ref{fig_spec_heat}.
The maximum of the curve is close to the transition temperature
determined above. It seems that the magnitude of the specific heat is 
independent of the system size (for large enough systems), which implies 
a critical exponent of $\alpha\approx 0$, where $\alpha$ is defined by 
$C\propto|T-\Tc|^{-\alpha}$ close to the critical point.
This is in agreement with the scaling relation $\alpha+2\beta+\gamma=2$
(Rushbrooke's law).\cite{new99}

For comparison, the values obtained by Jeong and Steinhardt\cite{jeo93} 
for stackings of Penrose rhombus tilings are $\beta=0.2$,
$\gamma=1.6$, $\nu=1.6$, and $\alpha=0$. These values are in good
agreement with our result (except for the exponent $\beta$, whose
uncertainty is relatively large in our case).

\section{Some further remarks}
\label{sec_remarks}

\begin{figure}[b]
\includegraphics[width=\columnwidth]{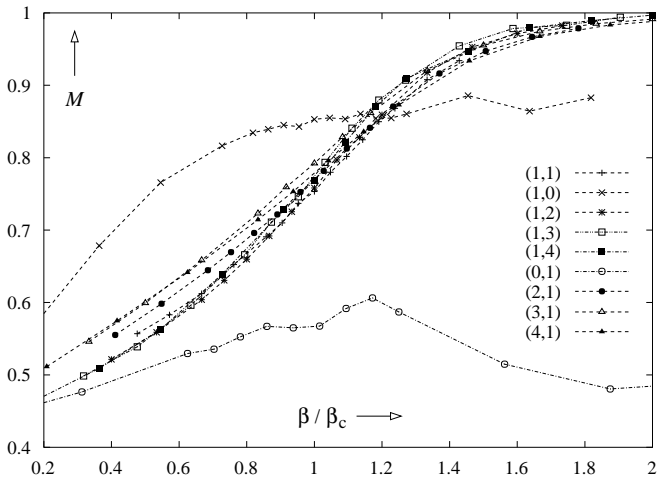}
\caption{Behaviour of the sheet magnetization for different ratios of
intralayer and interlayer coupling strength. All curves are for the
same approximant. The case of equal coupling strengths is labeled 
by (1,1), the case where the intralayer coupling is twice the interlayer 
coupling is labeled by (2,1), and so on.}
\label{fig_energieparam_magnet}
\end{figure}

To test the reliability of our results, we have run simulations
with different ratios of the intralayer and interlayer coupling strength.
It is evident that this influences the value of the critical
temperature and hence leads to different temperature
scales. Normalizing the temperature to the particular critical point,
i.~e., using $\beta/\betac$ as temperature scale, we obtain the same
behavior of the magnetization, independent of the relative coupling
strengths (Fig.~\ref{fig_energieparam_magnet}). 
Only the curves where one of the coupling energies is set to zero
show a different behavior, especially when the energy for the cluster
coupling is zero, labeled by (0,1). In this case, there is no interaction
inside the layers, so the order parameter is minimal, i.~e., zero for
infinite system size (see discussion of Fig.~\ref{fig_magnet} in
Sec.~\ref{sec_phasetrans}).
On the other hand, in the case (1,0) where the coupling energy in 
stacking direction is zero, there is still some kind of purely 
geometrical coupling between the layers due to the flip constraint
(Fig.~\ref{fig_flip-constraint}). This yields an order parameter
larger than zero. However, it does not approach $M=1$ at high $\beta$ (low
temperatures) since this geometrical coupling only restricts the
number of possible flips but does not enforce congruent layers.

\begin{figure}[t]
\includegraphics[width=\columnwidth]{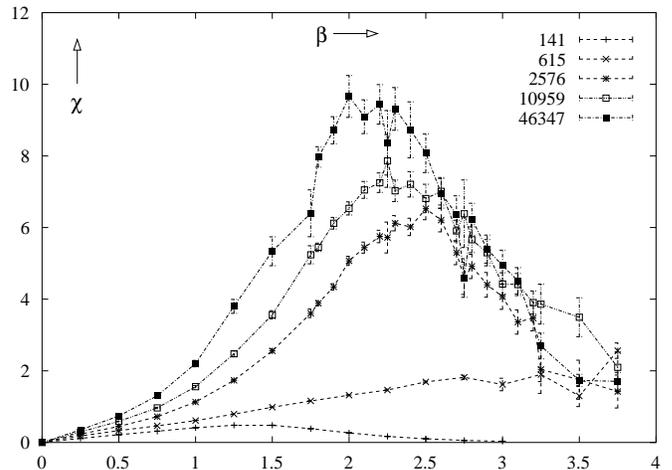}
\caption{System size dependence of the susceptibility vs.~inverse
temperature in the case of the relaxed rule.}
\label{fig_suscept_relax}
\end{figure}

We have also studied a 3D version of the {\it cluster density maximization} 
model obeying the {\it relaxed} rule (Sec.~\ref{sec_2d-model}). 
In this case, the intralayer energy is just the negative of the number
of clusters. (For the interaction in stacking direction we keep the
coupling described in Sec.~\ref{sec_3d-model}). Each ground state now 
consists of supertile random PPTs in each layer, which are all congruent. 
These supertile random PPTs are locally ordered, but show disorder at 
larger scales in terms of disoriented A-overlaps which violate the 
{\it perfect} overlap rule. The question is whether this partial,
local order distinguishes the ground state sufficiently from the 
disordered high-temperature state, so that a phase transition between 
the two can take place.
We cannot answer this question yet, because 
the curves obtained for the magnetization are too ``noisy'' 
(although the error bars are small), and the behavior of the 
susceptibility (Fig.~\ref{fig_suscept_relax}) does not allow to 
decide whether or not it diverges at the ``critical point'', at least not
for the accessible system sizes. It could well be that the dynamics is
confined to subsets of the phase space which are separated by high energy 
barriers from other such subsets, thus breaking the ergodicity of the 
simulation.

\section{Summary}
\label{sec_summary}

We have presented a cluster model for 3D decagonal quasicrystals which
shows a continuous phase transition at finite temperature from the
ordered low-temperature state to the disordered high-temperature
state. Using the sheet magnetization as order parameter and its
associated susceptibility, the critical exponents of this transition 
have been determined. Within the statistical errors, these critical 
exponents are in good agreement with the values obtained by 
Jeong and Steinhardt\cite{jeo93} for stackings of Penrose rhombus 
tilings (for comparison, see Table~\ref{tab_crit_exp}). 

\begin{table}[t]
\caption{Comparison of the critical exponents}
\begin{ruledtabular}
\begin{tabular}{lllll}
& $\alpha$ & $\beta$ & $\gamma$ & $\nu$ \\
\hline
{\bf Cluster model} & {\bf 0.0} & {\bf 0.08} & {\bf 1.7} & {\bf 1.6} \\
Tiling model\cite{jeo93} & 0.0 & 0.2 & 1.6 & 1.6 \\
2D Ising model\cite{ons44} & 0 & 0.125 & 1.75 & 1 \\
3D Ising model\cite{fer91} & 0.11 & 0.33 & 1.24 & 0.63 \\
\end{tabular}
\end{ruledtabular}
\label{tab_crit_exp}
\end{table}

The notion of spin sheets suggests a certain resemblance of our model
to the 2D Ising model. Indeed, there is some similarity also in the 
critical exponents (Table~\ref{tab_crit_exp}). We have to point out, 
however, that there is an essential difference to the 2D Ising model. 
The interaction between the clusters inside the layers is rather
different from the interaction between the layers in the stacking
direction, i.~e., the interactions in the system are highly anisotropic.
Furthermore, in the Ising model only next-neighbor interactions are
considered, whereas in our cluster model (and also in the Penrose
tiling) the coupling between the clusters (or the coupling which
arises from the matching rules in the Penrose tiling, respectively)
are of longer range. Hence, it is not too surprising that the critical
exponents of the 2D Ising model deviate from the ones obtained for 
our model (and those obtained by Jeong and Steinhardt\cite{jeo93}).
Nevertheless, one can say that the critical exponents for the 3D 
decagonal quasicrystals resemble those of the 2D Ising model more 
than those of the 3D Ising model.

\end{document}